\newcommand{\DyT}{Dy$_{2}$Ti$_{2}$O$_{7}$}

\newcommand{\HoT}{Ho$_{2}$Ti$_{2}$O$_{7}$}

\documentclass[12pt]{article}
\usepackage{graphicx}
\usepackage{setspace}
\usepackage{color}
\usepackage[super,sort&compress]{natbib}
\usepackage{multirow}
\usepackage{bm}
\usepackage{amsfonts}
\usepackage{amssymb}
\usepackage{enumerate,ulem}
\usepackage{float}
\usepackage{ulem} 
\usepackage{color}

\begin{document}

\begin{center}
{\bf \large Brownian Motion and Quantum Dynamics of Magnetic Monopoles in Spin Ice}
\vspace{0.5cm}

\vspace{1cm}

L. Bovo$^1$, J. A. Bloxsom$^1$, D. Prabhakaran$^2$, G. Aeppli$^1$, S. T. Bramwell$^1$.  

{\it 1. London Centre for Nanotechnology and Department of Physics and Astronomy, University College London, 17-19 Gordon Street, London, WC1H OAH, U.K.} \\
{\it 2. Department of Physics, Clarendon Laboratory, University of Oxford, Park Road, Oxford, OX1 3PU, U.K. }

\end{center}

{\bf Spin ice illustrates many unusual magnetic properties, including zero point entropy, emergent monopoles and a quasi liquid-gas transition. To reveal the quantum spin dynamics that underpin these phenomena is an experimental challenge. Here we show how crucial information is contained in the frequency dependence of the magnetic susceptibility and in its high frequency or adiabatic limit. These measures indicate that monopole diffusion is strictly Brownian but is underpinned by spin tunnelling and is influenced by collective monopole interactions. We also find evidence of driven monopole plasma oscillations in weak applied field, and unconventional critical behaviour in strong applied field. Our results resolve contradictions in the present understanding of spin ice, reveal unexpected physics and establish adiabatic susceptibility as a revealing characteristic of exotic spin systems.}

\newpage

\section{Introduction}

In spin ice materials like \HoT~or \DyT~\cite{Harris,BramwellHarris,Ramirez,denHertog,CMS,Ryzhkin,BramwellGingras} magnetic rare earth ions (e.g. Ho, Dy) occupy a lattice of corner-linked tetrahedra. In the low temperature spin ice state two atomic magnetic moments or `spins' point into, and two point out of each tetrahedron (Fig. 1). This is equivalent to the ice rule that determines proton configurations in water ice~\cite{Harris,BramwellHarris}, and hence spin ice has a residual entropy equal to the Pauling entropy of water ice~\cite{Ramirez}. The thermodynamic properties of spin ice are well described by a classical spin Hamiltonian with a dominant dipole-dipole interaction~\cite{denHertog,Yavorskii}. The self-screening of the latter establishes the ice rule ground state~\cite{denHertog}, but this property does not extend to excited states~\cite{CMS}. A spin flip out of the ice rule manifold creates a dipolar magnetic excitation that may fractionalise to produce free defects (Fig. 1). These inhabit the diamond lattice formed by tetrahedron centres and behave as magnetic monopoles on account of the integrated dipole-dipole interaction~\cite{CMS}.

The spin ices are part of the family of rare earth pyrochlores, a series of frustrated magnets for which collective quantum effects have been widely discussed~\cite{Champion,Stasiak,Zhitomirsky,Gingras,Mirebeau,Balents,Chang}. Recent theoretical work~\cite{Shannon,Benton} does not rule out the possibility that such effects may be relevant to \HoT~and \DyT, but to a good approximation, the monopoles may be treated as classical objects, with local quantum mechanics setting local parameters such as attempt frequencies. 

The magnetic monopole current density in spin ice is defined as the rate of change of magnetization: ${\bf J} = \partial {\bf M}/\partial t$, with the conductivity proportional to the monopole density~\cite{Ryzhkin}. However, even in an infinite system, magnetic monopoles in spin ice cannot sustain a direct current, on account of the destruction of the spin ice entropy by magnetization of the system~\cite{Ryzhkin}. This means that direct current (dc) `magnetricity' in spin ice~\cite{BramwellGiblin,Giblin} is necessarily transient~\cite{Jaubert,Jaubert2}. Alternating current (ac) magnetricity does not suffer from this limitation as monopoles can in principle be driven indefinitely back and forth by an oscillating magnetic field. The theory of ac-current~\cite{Ryzhkin} has not yet been tested as existing ac-magnetization studies either precede the theory\cite{Snyder,Matsuhira} or focus on the low temperature regime~\cite{Mats-2,Quilliam} where complicating factors are expected~\cite{Jaubert,Jaubert2,Giblin}. In Section 2 we present the first experimental test of the theory of Ref. \cite{Ryzhkin}, where we rigorously confirm a number of ideas and arguments about monopole diffusion~\cite{Jaubert,Jaubert2,CMS2} and spin tunnelling~\cite{Ehlers,Ehtwo,Snyder}, and derive new information on the microscopic processes involved. 

As shown in Fig. 2, our dynamical magnetization measurements also estimate the isothermal susceptibility $\chi_T$ and the adiabatic susceptibility $\chi_S$. While the former is a much discussed magnetic response function, the latter is typically neglected. Nevertheless, our experimental data clearly show (Fig. 2) that $\chi_S$ is finite, with the ratio $\chi_S/\chi_T$ increasing with increasing applied static magnetic field.
In Section 3 we report a striking contrast between the temperature dependence of $\chi_S$ and $\chi_T$ in weak applied field, showing that $\chi_T$ is best interpreted as a spin response, while $\chi_S$ is best interpreted as a monopole response. This contrast has its root in the fact that configurational entropy ultimately confines the monopoles when they are driven by a magnetic field~\cite{Ryzhkin}. 

In strong applied field along the cubic [111] direction, spin ice exhibits a liquid-gas type phase transition with a critical point~\cite{Sakakibara} at $\mu_0 H_C= 0.929$ $T$, $T_C= 0.36$ $K$. This transition has been interpreted as a monopole condensation~\cite{CMS} and has been treated in renormalisation group theory~\cite{Shtyk}. In Section 4 we extend our comparison of $\chi_T$ and $\chi_S$ to the `supercritical regime' at $T> T_C$, where we observe strong signatures of critical behaviour and find that monopoles behave increasingly like dipole pairs, in agreement with comments of Ref. \cite{Shtyk}. 
 
The temperature and field regimes probed in this paper are illustrated in Fig. 1, where we broadly define regimes of monopolar response and dipolar response. It should be emphasised that, as in other cases where novel quasiparticles accurately account for the low-energy physics, monopole and spin descriptions are never in conflict. Instead, certain properties are best discussed in terms of spins and others are best discussed in terms of monopoles. One result of our study is to clarify how this division should be made. 

\section{Magnetic Relaxation in Zero DC-Field}

According to Ref.\cite{Ryzhkin} (see also Ref. \cite{STB}) the frequency ($\omega$) dependent susceptibility arising from monopole currents should be described by the following equation with $\alpha =\chi_S = 0$ and $\tau$ a relaxation time:  
\begin{equation}\label{tau}
\frac{\chi({\omega})-\chi_S}{\chi_T-\chi_S} = \frac{1}{1+ (i \omega \tau)^{1-\alpha}}.
\end{equation}
With finite $\chi_S$ and parameter $\alpha$ this expression coincides with those of Casimir - Du-Pr\'e~\cite{Casimir} and Cole - Cole~\cite{Cole} for ordinary magnetic relaxation, but there is an important difference: here it is simplest to understand the behaviour of $\tau$ in terms of monopole dynamics rather than interacting dipoles. In particular the relaxation rate may be written $\tau^{-1} = \mu_0uQ x/V_0$ where $u$ is the monopole mobility, $Q$ the monopole charge, $x$ the total monopole density per diamond lattice site, and $V_0$ the volume per lattice site~\cite{STB}. The density $x(T)$ evolves with temperature in a way that cannot be expressed in closed form~\cite{CMS2}. However, we could measure $x(T)$ by fitting specific heat data to Debye-H\"uckel theory~\cite{CMS2,Zhou} so that by dividing our measured $\tau$ by $x(T)$ we were able to derive the mobility $u(T)$ as a function of temperature (Methods). Representative experimental data and results, along with characteristic fits, are shown in Fig. 3. 

Referring to Fig. 3, the model fits the experimental data well at $T > 3. 5$ K but describes only the high frequency part at the lowest temperatures (Supplementary Information, Fig. S1). At $T> 10$ K the apparent mobility diverges in accord with an expected Orbach type spin flip process~\cite{Ehlers,Ehtwo,Snyder} that is not considered further here. At lower temperatures $u(T)$ becomes accurately proportional to $1/T$ which is consistent with the Nernst-Einstein-Smoluchowski equation for the Brownian diffusion of magnetic monopoles: 
\begin{equation}
u = \frac{DQ}{kT}.
\end{equation}  
Here, the diffusion constant $D$ is temperature-independent, as shown in the inset of Fig. 3. The athermal diffusion constant shows that the observed temperature dependence of the magnetic relaxation\cite{Snyder}, in this temperature range, is completely accounted for by the temperature evolution of the monopole density and the temperature factor characteristic of a diffusive process (Supplementary Information, Fig. S2). This general behaviour is insensitive to small applied field (typically $\mu_0 H < 50$ ${\rm mT}$, Fig. 3). Writing $D= \nu_0a^2/6$ where $a$ is the diamond lattice constant and $\nu_0$ the monopole hop rate~\cite{CMS2,STB}, we find a temperature independent hop rate of $\nu_0 =2.43(6) \times 10^3$ s$^{-1}$. This athermal hop rate may be treated as evidence of quantum tunnelling of the spin involved in the monopole hop (Fig. 1).  

Our results are fully consistent with the theory of Ref.~\cite{Ryzhkin} and the numerical analysis of Ref.\cite{Jaubert} but also indicate an essential refinement that must be made. That is, we find a finite $\alpha$, suggesting a significant dispersion of relaxation times, as previously observed~\cite{Matsuhira}, rather than the single relaxation time assumed in the theory. However, the theory neglects monopole interactions which might be expected to influence the hopping rate of individual monopoles. To capture this we assume that in zero applied field, spins are flipped by transverse fields~\cite{Rosenbaum} arising from the dense ensemble of atomic dipoles, and we decompose the instantaneous local transverse dipolar field as follows: 
\begin{equation}\label{field}
 H  = H_0( 1+  h_1 + h_2\sqrt{x}),
\end{equation}
where $H_0$ is an effective field that causes flipping at rate $\nu_0$,  $h_1$ determines a static dispersion in the latter field, and $h_2 \sqrt{x}$ determines the local field arising from the monopole gas, which at the level of Debye-H\"uckel theory~\cite{STB} scales as $\sqrt{x}$. Assuming uncorrelated contributions it may be shown (Methods) that the variance of the logarithmic relaxation time is given by:
\begin{equation}\label{sig}
\sigma^2_{\ln \tau} = \sigma^2_1 + x \sigma_2^2, 
\end{equation}  
where the subscripts 1 and 2 refer to the fields $h_1$ and $h_2$, respectively. Our measured $\alpha(T)$ may be transformed~\cite{Zorn} to give the quantity on the left (Methods) and hence we can test the above expression. Fig. 4 confirms a very satisfactory agreement between theory and experiment in zero and weak applied field, with the fitted $\sigma_2$ increasing rapidly in an applied dc field of 1 mT, but thereafter more slowly. Possible ambiguities in our interpretation are discussed in the Supplementary Information (Fig. S3), but at the very least we may conclude that observed dispersion of rates is in large part a monopole property. 

\section{Isothermal and Adiabatic Susceptibilities}

The isothermal susceptibility $\chi_T$ extracted from the fits to theory is in close agreement with the directly measured $\chi_T$ (Supplementary Information, Fig. S4).
In theory the isothermal susceptibility is twice the Curie susceptibility~\cite{Ryzhkin}, $\chi_T = 2C/T$, but recent work~\cite{Jaubert-TSF} has established that in spin ice there is a crossover from a Curie constant $C$ at very high temperature to the expected $2C$ at low temperature. Our results indicate that $C\approx4.25$ in the temperature range explored (Supplementary Information, Fig. S4) which we interpret as evidence for this crossover: however a much more detailed study of the Curie Law crossover in ${\rm Dy_2Ti_2O_7}$ would be worthwhile. The Curie like $\chi_T$ is of course characteristic of a spin system: indeed there is no direct monopole signature in this quantity. This may be traced to the configurational entropy in the problem, which in applied field confines the monopoles~\cite{Ryzhkin,STB}, making the magnetic response spin-like at long time.

The thermodynamic adiabatic susceptibility $\chi_S$ is the ac-susceptibility extrapolated to infinite frequency (or more strictly to a frequency where spin-spin relaxation is active but where spin-lattice relaxation is not~\cite{Casimir}). Fig. 5 (bottom) illustrates a striking correlation between our measured adiabatic susceptibility $\chi_S(T)$ and the measured monopole density $x(T)$ (Supplementary Information, Fig. S5 and S6, for further discussion). 
Thus we find $\chi_S = \chi_0 x(T)$ with $\chi_0 = 0.030(1)$, a temperature-independent constant. In a monopole picture, we may imagine a frictionless, and hence reversible, displacement of magnetic monopoles by distance $r$ in the applied field - like a driven plasma oscillation. If we write the force on a positive monopole as $\mu_0 H(\omega) Q = K r(\omega)$, where $K$ is the force constant, then the magnetization is $M(\omega) = (x/V_0) Q r(\omega)$, from which $\chi_S = x \mu_0Q^2/KV_0$, as observed. From the value of $\chi_S$, we find $K \approx 0.12$ $Nm^{-1}$, implying an energy barrier between lattice sites at a distance $r=a/2$ of order $200$ $K$. The latter seems too large to be a Coulombic barrier, and is more likely connected with the crystal field energy scale of several hundred kelvin. The frictionless oscillation of the monopole ensemble is reminiscent of a plasma oscillation in an electrical plasma, though the absence of an accelerative term in the equation of motion means that the monopole plasma oscillation cannot occur in the absence of a driving field. 
Of course a finite $\chi_S$ in a magnetic system can always be formally represented as an oscillation of magnetic charge, but in this case our result shows it to be associated with the motion of recognisable positive and negative magnetic monopoles.

In a magnetic system $\chi_S$ is always less than the isothermal susceptibility $\chi_T$, as it obeys the thermodynamic relation: 
\begin{equation}\label{thermo}
\chi_S = \chi_T - \frac{T \left(\partial M/\partial T)^2\right)}{C_H},
\end{equation}
where $C_H \ge 0$ is the specific heat at fixed applied field $H$. For a paramagnetic rare earth salt a typical behaviour of $\chi_S$ would be to roughly track the increase of $\chi_T$ as $T \rightarrow 0$ according to the Curie Law $\chi_T = C/T$. The striking difference we observe between $\chi_T(T)$ and $\chi_S(T)$ (Fig. 5) reflects transition from spins to monopoles as the natural variables by which to describe the magnetic response, monopoles being more appropriate at high frequency. 

Nevertheless, we can explore the origin of $\chi_S$ in spin language if we consider a monopole as a label for a set of `flippable' spins (Fig. 1). We assume that the adiabatic susceptibility is equal to the isolated susceptibility, which in a semi-classical approximation is given by~\cite{Broer}:
\begin{equation}
\chi_S =  Z^{-1} \sum_n \left(\frac{\partial M_n}{\partial H}\right) e^{-E_n/kT}.
\end{equation}
Here $Z$ is the partition function, $n$ labels the energy states of the system, and $M_n$ is the magnetic moment per unit volume of the state $n$. If the ground state is assigned null moment and the monopole excited state is assigned $\partial M_n/\partial H = \chi_0$, where $\chi_0$ is temperature independent, then we obtain our experimental observation that $\chi_S = \chi_0 x(T)$. Since $M_n = V^{-1} \partial E_n/\partial (\mu_0 H)$ (where $V$ is volume) our result reveals a quadratic term in the energy per monopole: $E'_n = (V \mu_0\chi_0/2)H^2$.  

A quadratic energy expression generally indicates `stretchable' magnetic moments. A small quadratic (Van Vleck) term is expected for a free Dy$^{3+}$ ion through mixing of the ground state with states of higher total angular momentum $J$. However, in our case, the observation that only flippable spins contribute to $\chi_0$ and that flippable spins and non-flippable ones are distinguished only by a thermal energy scale at these temperatures, appears to rule out any single spin mechanism. It is interesting to note that the monopole spin texture is predicted to produce an electric dipole~\cite{Khomskii} and it appears from our result that it is associated with a magnetic polarisability as well.

These findings have very important ramifications for the monopole description of spin ice. Previous work by neutron spin echo\cite{Ehlers,Ehtwo} on \HoT ~and $\mu$SR~\cite{Lago} on \DyT ~has suggested a high frequency response that thermally evolves to low temperature, and that at first sight seems disconnected from the monopole picture. However, our results indicate that the dynamical spectrum in the approach to the high frequency limit is fully accounted for by magnetic monopoles, and they clearly explain the thermal evolution observed in the previous work. A very recent thermal conductivity study\cite{Kolland} indirectly estimates a diffusion constant for magnetic monopoles that is much faster than ours, but our adiabatic susceptibility results show that there is no necessary contradiction, as monopoles mediate a dynamical response over a very broad frequency range. Finally our results rule out any significant spectral weight beyond that associated with monopoles, contrary to a recent proposal~\cite{Dunsiger}. 

\section{Adiabatic Susceptibility in Applied Field}

Fig. 1 shows the spin ice phase diagram for a dc-magnetic field applied along the cubic [111] direction. A small applied field orders one spin per tetrahedron in the pyrochlore structure, but maintains the ice rule of two spins in and two out per tetrahedron, thus creating the so called `kagome ice' phase of two-dimensional disordered sheets, which still possess residual entropy~\cite{Mats-kag,Sakakibara,Hiroi}. With increasing field at $T=0$ there is a breaking of the ice rules, pictured as the flipping of one spin per tetrahedron, to create an ordered `three in, one out' state. Extending from this point a first order phase transitions that terminates in a critical end point. The positive slope of this line reflects the destruction of the spin ice entropy by the applied field, according to the Clapeyron equation. In the monopole representation, the applied field tunes the chemical potential of monopole-antimonopole pairs such that the increased monopole density drives a first order condensation from a sparse monopole fluid to a dense `liquid' (or perhaps better, ionic crystal) of alternating positive and negative monopoles~\cite{CMS}. The detailed theory of magnetic relaxation near the critical point~\cite{Shtyk} predicts mean field critical exponents modified by logarithmic corrections. Here we are interested in the supercritical region at temperatures well above the critical point, where the system may be described as a dense monopole plasma. Recently, a peak in the ac-susceptibility at finite frequency was observed in this region (Matthews, M. \& Schiffer P. unpublished). We examined the behaviour of the adiabatic susceptibility as a function of field in this regime, to compare it with our zero field measurement. 

In weak fields ($\mu_0 H \lesssim 0.3$ $T$) the thermal evolution of $\chi_S(T)$ shows a slow increase with field, including a noticeable peak at higher temperature (Fig. 6). In much stronger fields ($\mu_0 H \gg1$ $T$) the adiabatic response is completely suppressed as would be expected (Fig. 6), but at an intermediate field ($\mu_0 H\approx 0.920(8)$ $T$), $\chi_S(H)$ exhibits a striking peak very near to the (internal) field of the zero temperature phase transition. At this field the ice rule is locally broken~\cite{Mats-kag} and $1/4$ of the spins in the sample may then be flipped at zero energy cost. However, in contrast to the zero field result, $\chi_S(T)$ measured near this crossover field ($0.86$ $T$) exhibits a simple Curie law, $\chi_S = C'/T$ (Fig. 5), indicating a different type of magnetic current to that observed in the weak field limit, as anticipated in Ref.\cite{Shtyk}.
We may regard the magnetic response in this regime as characteristic of switching magnetic dipoles, rather than magnetic monopoles. Note that the temperature evolution was measured at this point just off the peak maximum as it was found that systematic errors in fitting to the Cole-Cole function are minimised at this point (Supplementary Information, Section 5). The Curie constant $C'$ may be calculated under the assumption that $1/4$ of the spins are thermally active and that these have a projection of $1/3$ of their full classical value on the field direction. Thus we predict $C' = C/36 = 0.1097$ where $C \approx 3.95$ is the high temperature Curie constant. A fit of the experimental data to the expression $\chi_S(\mu_0 H = 0.86$ ${\rm T}$)$ = a/(T-T_C)$ gave $a = 0.090(5), T_C = 0.4(2)$ in close agreement with our prediction (Fig. 5). 

The striking $1/T$ divergence and location of the peak position in the $H-T$ plane (Fig. 6) suggests that the adiabatic susceptibility is dominated by the classical critical point for monopole condensation (Fig. 1), for which the isothermal  susceptibility $\chi_T$ is predicted~\cite{Shtyk} to diverge as $1/|T-T_C|^{\gamma}$, with $\gamma=1$ and $T_C \ll T$ here. Usually the ratio $\chi_T/\chi_S$, analogous to the Landau-Placzek ratio in a fluid, should diverge towards the critical point. However, our data for $\chi_T$ and $\chi_S$ (Fig. 5) illustrate a fairly typical paramagnetic response, as discussed above, with $\chi_S \sim \chi_T$. Hence both $\chi_S$ and $\chi_T$ diverge, but the latter is always larger, as required by thermodynamics (Eqn. \ref{thermo}). 

According to Ref. \cite{Shtyk}, the field dependence of the susceptibility should be mean field like, with logarithmic corrections: hence we would expect $\chi \sim |H-H_C|^{(1/\delta) -1} $ with $\delta = 3$. However we observe an exponent of $2$ rather than $2/3$ (Supplementary Information, Fig. S7). Thus, defining reduced variables $t=T-T_C$ (with $T_C = 0.36$ here), and $h=H-H_C$, we find to a good approximation (suppressing dimensional constants):
\begin{equation}
\chi_S \sim \frac{1}{t + h^2}.
\end{equation}
This implies $h/\sqrt{t}$ scaling (Fig. 6, inset), which is formally characteristic of a zero dimensional phase transition
(Supplementary Information, Section 6). An alternative interpretation of the field dependence is in terms of a classical single spin flip process, associated with the `free' moments in the eventual ordered structure, which would be characterised by a response of the type $\chi_S \sim t/(t^2 + h^2)$ and hence $h/t$ scaling, but our data appears to distinctly rule against this possibility (Fig. 6, inset). Thus the behaviour of $\chi_S(H)$ seems inconsistent with both the `monopole' and $T= \infty$ fixed points. One possibility, as discussed further in the Supplementary Information (Section 6), is that the susceptibility is dominated by a zero temperature quantum critical point, but again with anomalous exponents. It is noteworthy that there have been reported several other examples of anomalous exponents in the quantum critical behaviour of rare earth magnets~\cite{Gabriel, LiEr}. 


\section{Conclusion}

While the concept of magnetic monopoles in spin ice is supported by much experimental evidence~\cite{Fennell, Morris,Kadowaki}, the microscopic mechanism of monopole motion has yet to be identified. Our investigation of \DyT~has isolated the characteristics of this mechanism to which any future theory must conform. Our firm result is that monopoles obey the Nernst-Einstein-Smoluchowski equation with temperature independent diffusion constant, a strong signature of Brownian diffusion. It should be emphasised that this is an experimental result and not a theoretical input.  

It is interesting to discuss these results in the context of band theory. Just as water ice can be thought of as an intrinsic protonic semiconductor~\cite{Pujol}, so spin ice can be thought of as an intrinsic semiconductor for magnetic monopoles. These are produced by the thermal unbinding (or `fractionalisation'~\cite{CMS}) of conventional magnetic excitons. They tunnel from site to site and have an effective mass determined by the inverse Debye length (proportional~\cite{STB} to $\sqrt{x/T}$). The fact that the carriers  support a temperature independent diffusion constant places an essential constraint on any theoretical development of this picture. 

We have shown that the adiabatic susceptibility gives a new perspective on the magnetic properties of spin ice, revealing a direct measure of the magnetic monopole concentration and critical behaviour in applied field. It would be useful to apply our methods to the low temperature regime, as zero-field measurements in that regime await an unambiguous interpretation in the monopole picture~\cite{Mats-2,Quilliam}, and the theory of Ref.~\cite{Shtyk} has yet to be comprehensively tested. More generally we may conclude that the adiabatic susceptibility, often ignored as an uninteresting by-product of ac-susceptibility analysis, may contain a wealth of information about strongly correlated spin systems at low temperature. 

Our results have revealed a new property of magnetic monopoles: their partial magnetic polarisation by an applied field. Taken together with the remarkable prediction that a monopole will carry an electric dipole moment (the equivalent of its spin, if we reverse the roles of electricity and magnetism)~\cite{Khomskii} a fascinating picture of the local properties of the monopole is starting to emerge. In general, the properties that we have discovered will have an important influence on any future application of magnetic monopoles in spin ice that seek to exploit their local quantum degrees of freedom.

\vspace{1cm}
{\bf Methods}

The dynamical magnetization of a $0.0326(1)$ $g$ cubic crystal of Dy$_2$Ti$_2$O$_7$ was measured with the ACMS (AC-Measurement System) option of a PPMS (Physical Property Measurement System, Quantum Design). Alternating and direct current magnetic fields ($H_{ac}$ and $H_{dc}$, respectively) were applied parallel to the cubic [111] axis of the sample. Data were collected at different temperatures between $1.9$ $K$ $ \le T \le 14$ $K$ in the ac-frequency range of $10$ $Hz$ to $10$ $kHz$. A variable dc field of $\mu_0 H = 0 - 10$ $T$ was applied (at low field the absolute field was calibrated in dc-sweep measurement). Scans were taken at different ac fields in the range $\mu_0 H_{ac} = 0.05 - 3\times 10^{-4}$ $T$ to dispel the possibility of non-linear response of the system. The results presented here were taken at $\mu_0 H_{ac} = 5 \times 10^{-5}$ $T$. Data were corrected taking into account a demagnetizing factor $\mathcal{D} = 1/3$ to give $\chi_{ac} = M_{ac}/\left((H_{ac} - \mathcal{D} M_{ac}\right)$. The calibrated response function of the instrument was checked by measurement of a very dilute paramagnetic salt (Supplementary Information, Fig. S5). 

The data were fitted to the phenomenological model for the frequency dependent susceptibility described in Ref.~\cite{Casimir,Cole}. By separating Eqn. \ref{tau} it is possible to derive analytical expressions for the real and imaginary parts and argand diagram (Cole-Cole plot), which were each fitted to the experimental data at a given temperature using a single set of parameters $\chi_S$, $\chi_T$, $\tau$, $\alpha$. 

The Cole-Cole formalism assumes a symmetric unimodal distribution of logarithmic relaxation times $\ln \tau'$ with mean $\ln(\tau)$. If the variance $\sigma^2_{\ln \tau'} \ll 1$ then it closely approximates $\sigma^2_{\nu}$ where $\nu=1/\tau'$ is the relaxation rate. It can be shown that~\cite{Zorn}:
\begin{equation}
\sigma^2_{\ln \tau'}  = \frac{\pi^2}{3}\left(\frac{1}{(1-\alpha)^2 - 1}\right). 
\end{equation}
In the text we label $\sigma^2_{\ln \tau'}$ simply as $\sigma^2_{\ln \tau}$ for ease of reading, although strictly $\tau$ is a fixed parameter at a given temperature. It should be noted that there is no general way to derive the true mean relaxation time $\langle \tau' \rangle$ from ac-susceptibility data: here we approximate it to the Cole-Cole parameter $\tau$. 

The dimensionless monopole density $x(T)$ was estimated by fitting experimental specific heat data to Debye-H\"uckel theory~\cite{CMS2,Zhou}. The specific heat was represented as the temperature derivative of the energy per diamond lattice site:
\begin{equation}
u = (-\mu+\mu_{DH})x+u_{DCM},
\end{equation}
where $\mu$ is the monopole chemical potential, $\mu_{DH}(T)$ is the Debye-H\"uckel correction, calculated self consistently with the dimensionless monopole density $x(T)$, and $u_{DCM}(T)$ is a correction for double charge monopoles. The experimental specific heat data, taken between $0.4$ $K$ $ \le T \le 10$ $K$, were fitted by adjusting $\mu$, with the best fit value $\mu/k = -4.33$ $K$. The theory is not exact in the temperature range of interest and Figs. 4 and 5 report an approximate envelope of systematic error, found by extrapolating the theory between low and high temperature according to different schemes (Supplementary Information, Section 7).  

To derive Eqn. \ref{sig}, assume $\tau^{-1} \propto H$, take logarithms and Taylor expand the right hand side of Eqn. \ref{field} to find 
$\ln{\tau} = h_1 + h_2\sqrt{x}$; then if $h_1$ and $h_2$ are uncorrelated, Eqn. \ref{sig} follows.

\vspace{1cm}
{\bf Supplementary Information}
Supplementary Information is linked to the online version of the paper 

\vspace{1cm}
{\bf Acknowledgements}

It is a pleasure to thank A. Fisher, C. Castelnovo and P. Holdsworth for valuable discussions; R. Aldus and M. Ellerby for their involvement in the specific heat work. The authors are also grateful to EPRSC for its support of the project.

\vspace{1cm}
{\bf Author Contributions}

LB did the experiment and analysed the data. STB conceived the project and derived the theory. STB, LB and GA planned the experiments and interpreted the results. STB and LB drafted the paper with input from GA. JAB derived the density versus temperature from specific heat data. DP grew the crystal. All authors commented on the final manuscript. 

\vspace{1cm}
{\bf Author Informations}

The authors declare no competing financial interests. Correspondence and requests for materials should be addressed to LB (l.bovo@ucl.ac.uk).

\newpage
\begin{figure}
\includegraphics[width=\columnwidth]{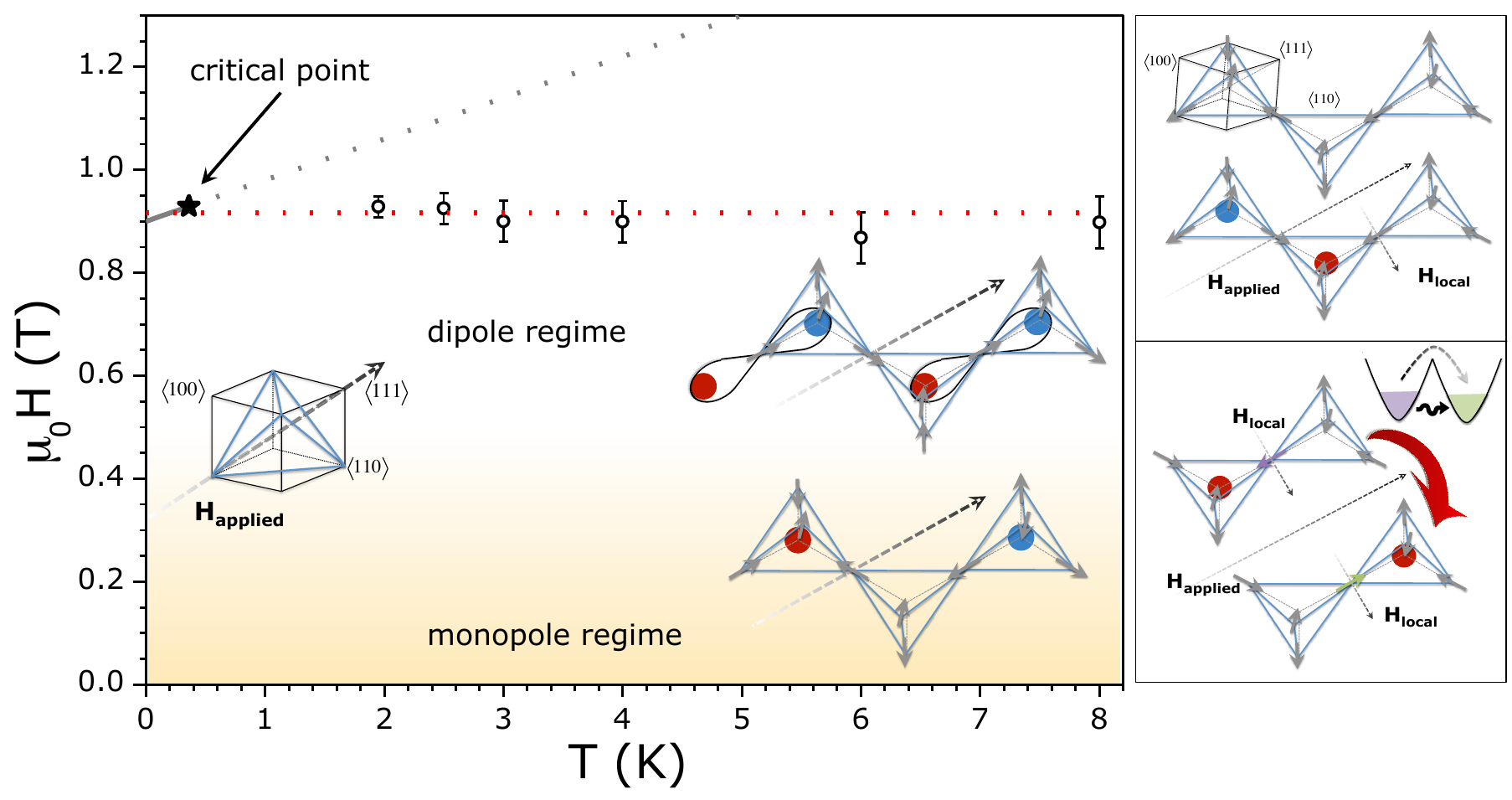} 
\caption{ {\bf Right: Fragment of the crystal structure of spin ice, illustrating the hopping of emergent magnetic monopoles.} Fragment of spin ice's cubic pyrochlore lattice, which consists of corner-linked tetrahedra, showing spin configurations (arrows). Top panel illustrates crystallographic axes, the applied field direction and how internal fields may be transverse to the local spin direction. Blue (red) circles represent negative (positive) monopoles. Bottom panel illustrates how a monopole hop can be associated with a spin flipped by a transverse field or a tunnelling event through a potential barrier.
{\bf Left: Temperature ($T$) versus field ($H \parallel [111]$) phase diagram of spin ice ${\rm Dy_2Ti_2O_7}$.} The full line is a line of first order phase transitions, terminating in a classical critical point, that has been interpreted as a monopole condensation~\cite{CMS}. Monopoles are deconfined in zero field but become confined in an applied field. The right hand diagrams show how the monopoles reform flippable spins or dipole pairs near the critical field. Dotted lines are guides to the eye. Experimental points with error bars show the applied field of the maximum in the adiabatic susceptibility measured here.}
\label{fig:ratio}
\end{figure}

\newpage
\begin{figure}
\includegraphics[width=\columnwidth]{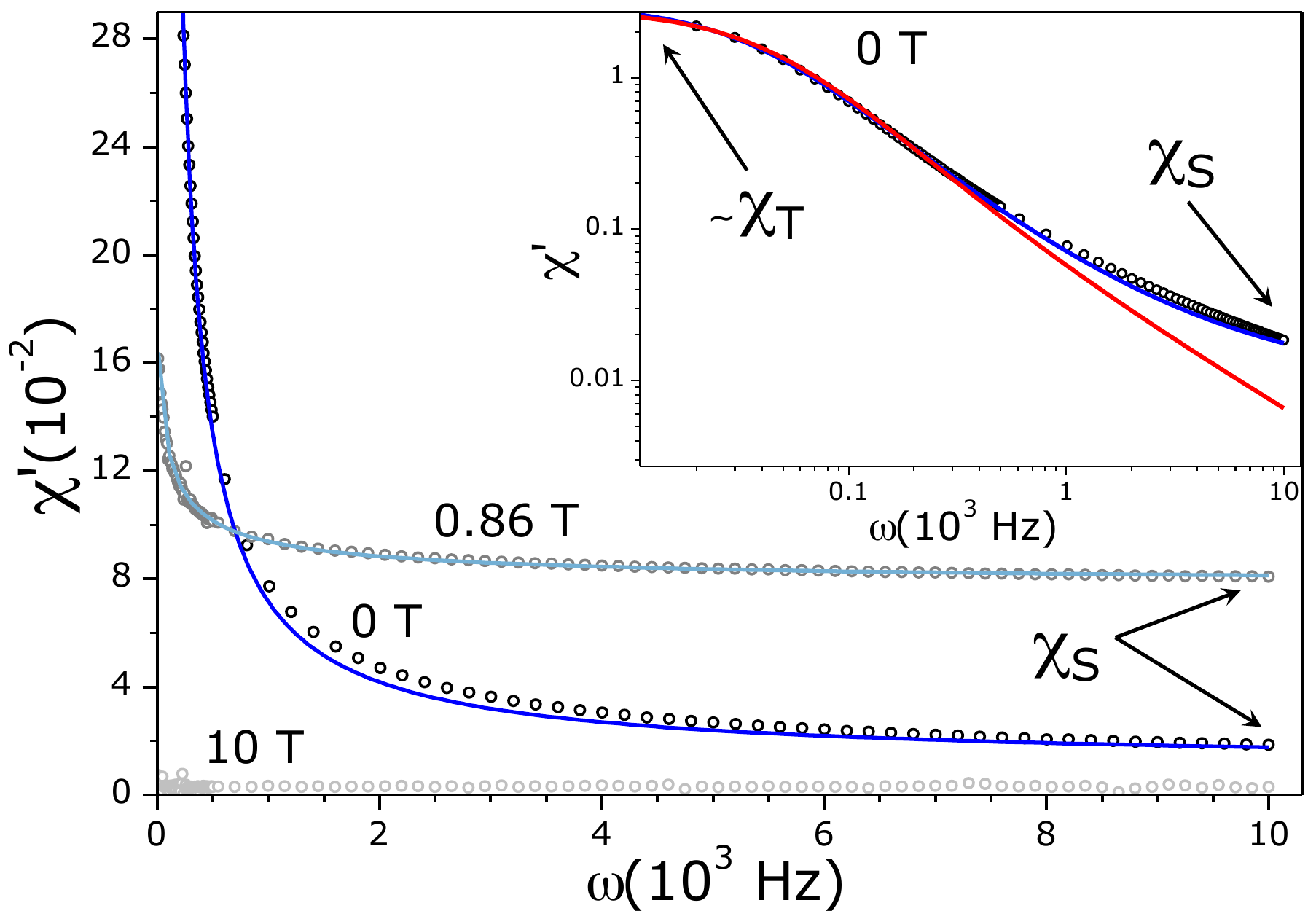} 
\caption{{\bf Experimental observation of a finite adiabatic susceptibility ($\chi_S$) and its relation to the isothermal susceptibility ($\chi_T$)}. These quantities are estimated as the real part of the frequency dependent susceptibility $\chi(\omega)$ in the limits $\omega \rightarrow \infty$ and $\omega \rightarrow 0$, respectively, as indicated in the main plot. Experimental data are at $T=1.95$ $K$ at applied static magnetic field $\mu_0 |H| = 0$ (black circles), $0.86$ (grey circles) and $10$ T (light grey circles). At 10 T both susceptibilities are nearly zero. At 0.86 T, the two are of similar magnitude. At zero applied field $\chi_S \ll \chi_T$, but still finite. The respective lines are the fit to a Cole-Cole model (see text). The inset shows the $\mu_0 H = 0$ data on log-log scales. The clear deviation from a linear curve at large $\omega$ confirms the presence of a finite offset, $\chi_S$. Here the blue line is the fit to the Cole-Cole function using finite $\chi_S$ and the red line is the same fit with $\chi_S$ constrained to be zero. }
 \label{fig:ratio}
 \end{figure}

\newpage
\begin{figure}
\includegraphics[width=\columnwidth]{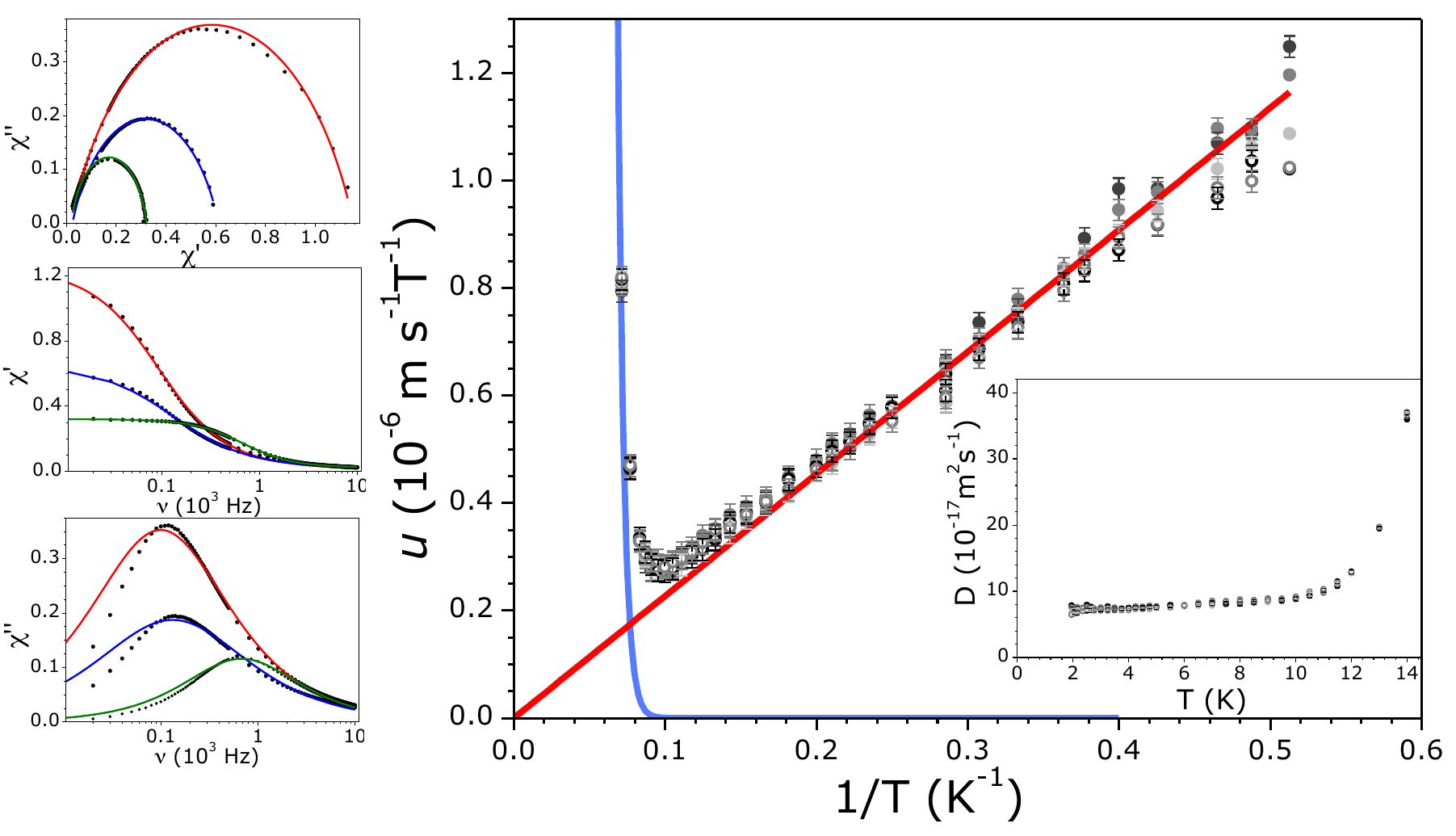} 
\caption{{\bf Ac-magnetricity and monopole diffusion in spin ice, ${\rm Dy_2Ti_2O_7}$.} Left: a representative fit to the ac-magnetization data at $H_{dc} = 0$: (middle) real and (lower) imaginary susceptibilities, and (upper) Cole-Cole plot (an argand diagram). The lines (red, T = 4.5 K; blue, T = 8 K; green, T = 14 K) fit all data with the same set of four parameters at a given temperature: $\chi_T$, $\chi_S$, $\alpha$ and $\tau$, as defined in the text. Right (main figure): monopole mobility measured at applied fields $\mu_0 |H| = 0$ (full black circles), 3 (full dark grey), 10 (full light grey), 18.5 (open black) and 38.5 (open grey) mT. The red line is $u = A/T$ (red line) where $A = 2.27(2)\times10^{-6}$ m K s$^{-1}$ T$^{-1}$. This is characteristic of Brownian diffusion of monopoles with a temperature-independent diffusion constant (inset). The blue line is $u = Be^{-C/T}$ (blue line) where $B = 39(1)$ m s$^{-1}$ T$^{-1}$, $C = 250(1)$ $K$, characteristic of a previously identified Orbach-like spin flip process arising from the coupling to excited crystal field states. This process is extinct below $10$ $K$, giving way to monopole diffusion as the cause of magnetic relaxation in spin ice. }
   \label{fig:ratio}
\end{figure}

\newpage
\begin{figure}
\includegraphics[width=0.9\columnwidth]{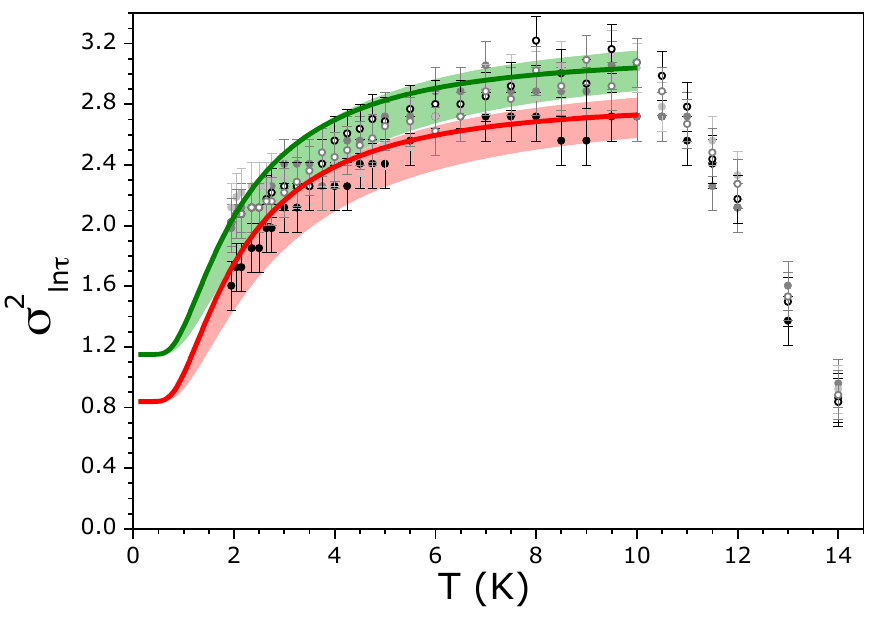} 
\caption{ {\bf Monopole signatures.} The experimental variance in logarithmic relaxation time $\sigma^2_{\ln \tau}$ (circles) compared with the predicted form for monopolar-field assisted tunnelling. Red and green indicate, respectively, applied fields of $\mu_0H = 0$ and the set of finite fields listed in the caption of Fig. 3. For each curve there are two fitted parameters $\sigma_1^2$ and $\sigma_2^2$, which describe the mean square static field and mean square monopole field, respectively. The line is the function $\sigma^2_1+ x \sigma^2_2$ where $x(T)$ is the monopole density. The shading indicates the maximum possible systematic error in the monopole density. We find $\sigma_{1,2} = 0.84(1), 3.40(5)$ respectively in zero field and $1.15(1), 3.40(5)$ in finite field. The deviation at $T > 10$ $K$ is related to a change in relaxation mechanism (see text).}
   \label{fig:ratio}
\end{figure}

\newpage
\begin{figure}
\includegraphics[width=0.6\columnwidth]{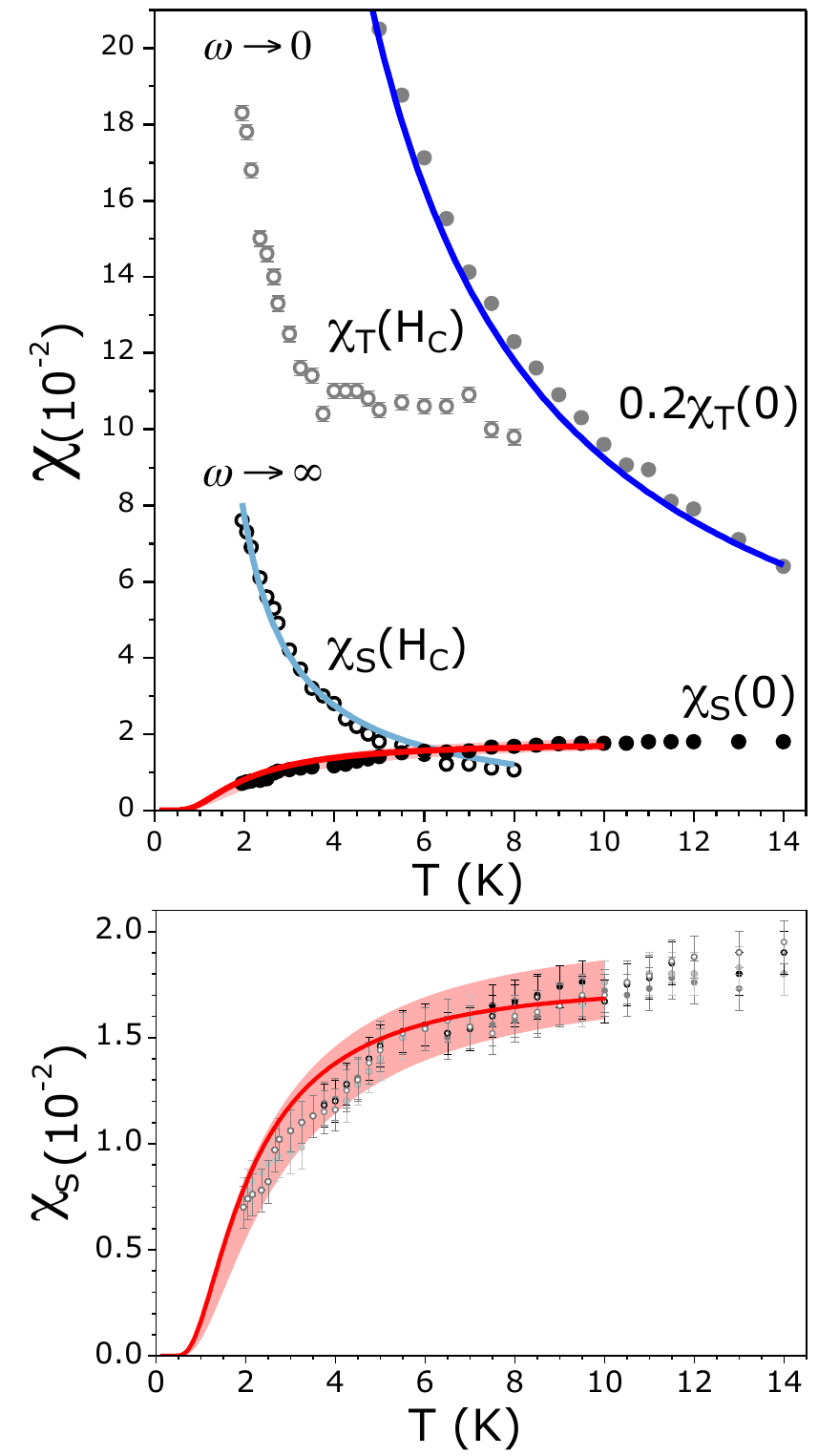} 
\caption{{\bf Monopole and dipole signatures revealed by comparing the measured adiabatic ($\chi_S$) and isothermal ($\chi_T$) susceptibilities.} 
Upper figure: the susceptibilities are shown as a function of applied  dc-field at zero field (black circles) and near the crossover field (grey circles) shown as a red dotted line in Fig. 1. The full red line is the experimentally measured monopole density fitted to $\chi_S(0)$ by the adjustment of a scale factor (see text). The blue line is a Curie-Weiss law fitted to $\chi_T(0)$ (Supplementary Information, Fig. S4). The light blue line is the theoretical prediction $\chi_S(H_C) = C'/(T-T_C)$ (see text).
Lower figure: Measured $\chi_S$ as a function of weak applied field $0\le \mu_0 H \le 0.05$ $T$ (circles, same colour code is maintained as in Fig. 3). The full red line is the experimentally measured monopole density fitted to $\chi_S(0)$ by the adjustment of a scale factor (see text), also showing the maximum systematic uncertainty in the measured monopole density (red shading).}
   \label{fig:ratio}
\end{figure}

\newpage
\begin{figure}
\includegraphics[width=\columnwidth]{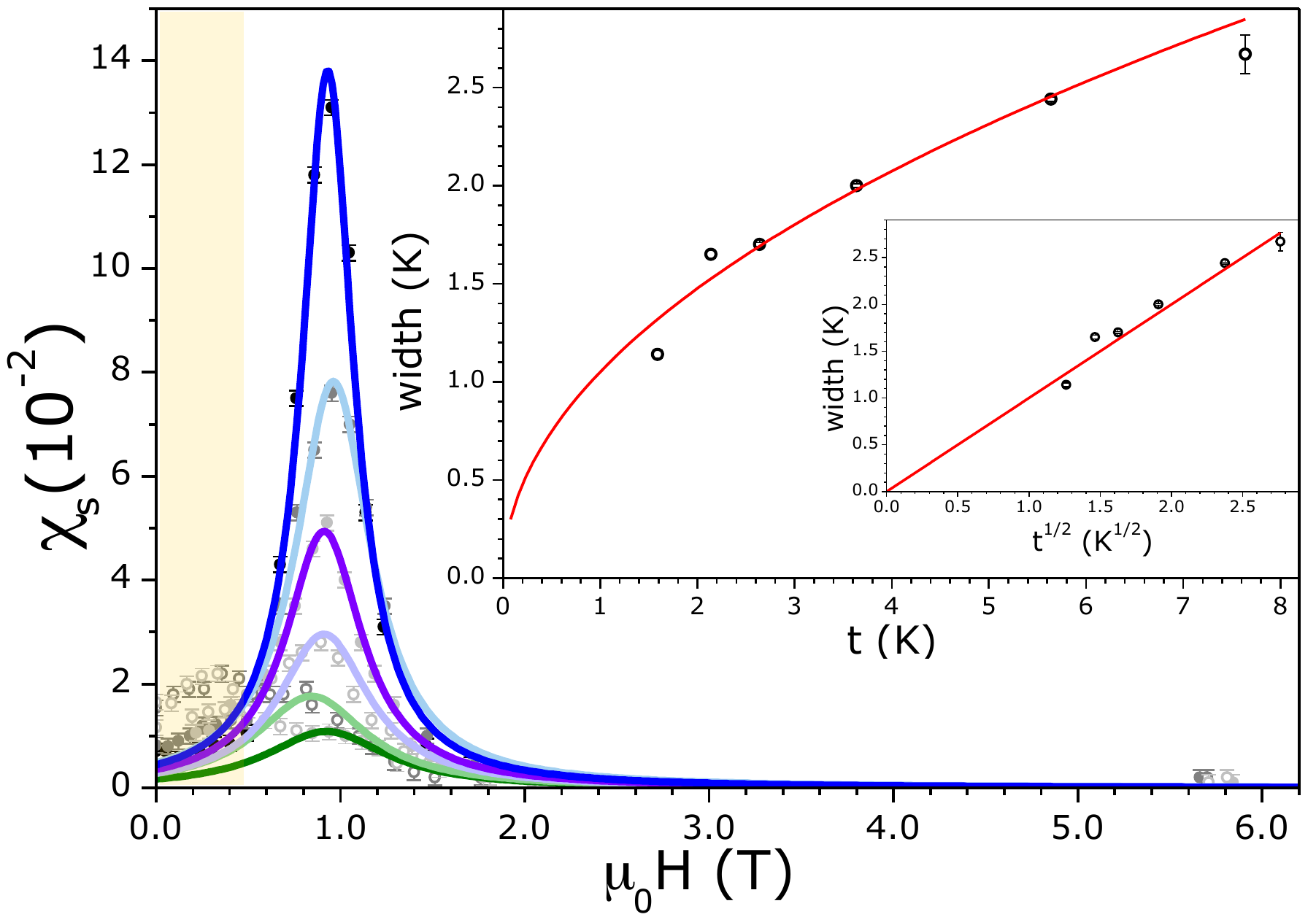} 
\caption{ {\bf Evidence of unconventional critical behaviour.} Main figure: adiabatic susceptibility versus applied magnetic field, showing an unusual Lorentzian field dependence, suggestive of anomalous critical behaviour. The figure illustrates zero response at strong field and a peak at the field where the ice rule breaks. The lines are fits to a Lorentzian function at the following temperatures: blue, $T = 1.95$ K; light blue, $T = 2.5$ K; purple, $T = 3.0$ K; lilac, $T = 4.0$ K; light green, $T = 6.0$ K; green, $T = 8.0$ K. Inset: scaling of Lorentzian width ($\Delta$): the red line is the function $\Delta \propto \sqrt{t}$. Experimental data appears to rule out the possibility of a different scaling - e.g. linear scale $h/t$.}
   \label{fig:ratio}
\end{figure}

\end{document}


\begin{center}
{\bf \large Brownian Motion and Quantum Dynamics of Magnetic Monopoles in Spin Ice: 
Supplementary Information}
\vspace{0.5cm}

\vspace{1cm}

L. Bovo$^1$, J. A. Bloxsom$^1$, D. Prabhakaran$^2$, G. Aeppli$^1$, S. T. Bramwell$^1$.  

{\it 1. London Centre for Nanotechnology and Department of Physics and Astronomy, University College London, 17-19 Gordon Street, London, WC1H OAH, U.K.} \\
{\it 2. Department of Physics, Clarendon Laboratory, University of Oxford, Park Road, Oxford, OX1 3PU, U.K. }

\end{center}

{{\bf 1. Model Used to Fit the data}

In the main text we note that the relaxation time $\tau$ does not have the same physical interpretation as that arising in conventional models of paramagnetic spin relaxation. However this fact may be neglected for fitting purposes, and the data analysis reduces to a familiar problem of fitting $\chi(\omega)$ to conventional phenomenological forms. 

The Debye model, with a single exponential relaxation of the susceptibility fails in the description of experimental data. Various empirical equations have been formulated to give the variation of ${\rm Im}[\chi(\omega)] = \chi''$ with ${\rm Re}[\chi(\omega)] = \chi'$ as the frequency is varied, taking into account the presence of different distributions of relaxation times. In particular, as discussed in the main text, we tested two of the most commonly used models: the Cole-Cole (CC)~\cite{Cole} and Cole-Davidson (CD)~\cite{DC} formalisms. Fig. S1 shows an illustrative comparison between the two representations. The CC formalism works rather well in all range of frequency for a wide interval of temperature, $3.5 \le T \le 14$ $K$. At low temperatures, in the low frequency regime, the experimental data start to deviate form the model, showing an increasing asymmetry. 
\begin{figure}
\renewcommand{\thefigure}{S\arabic{figure}}
\centering
\includegraphics[width=0.9\columnwidth]{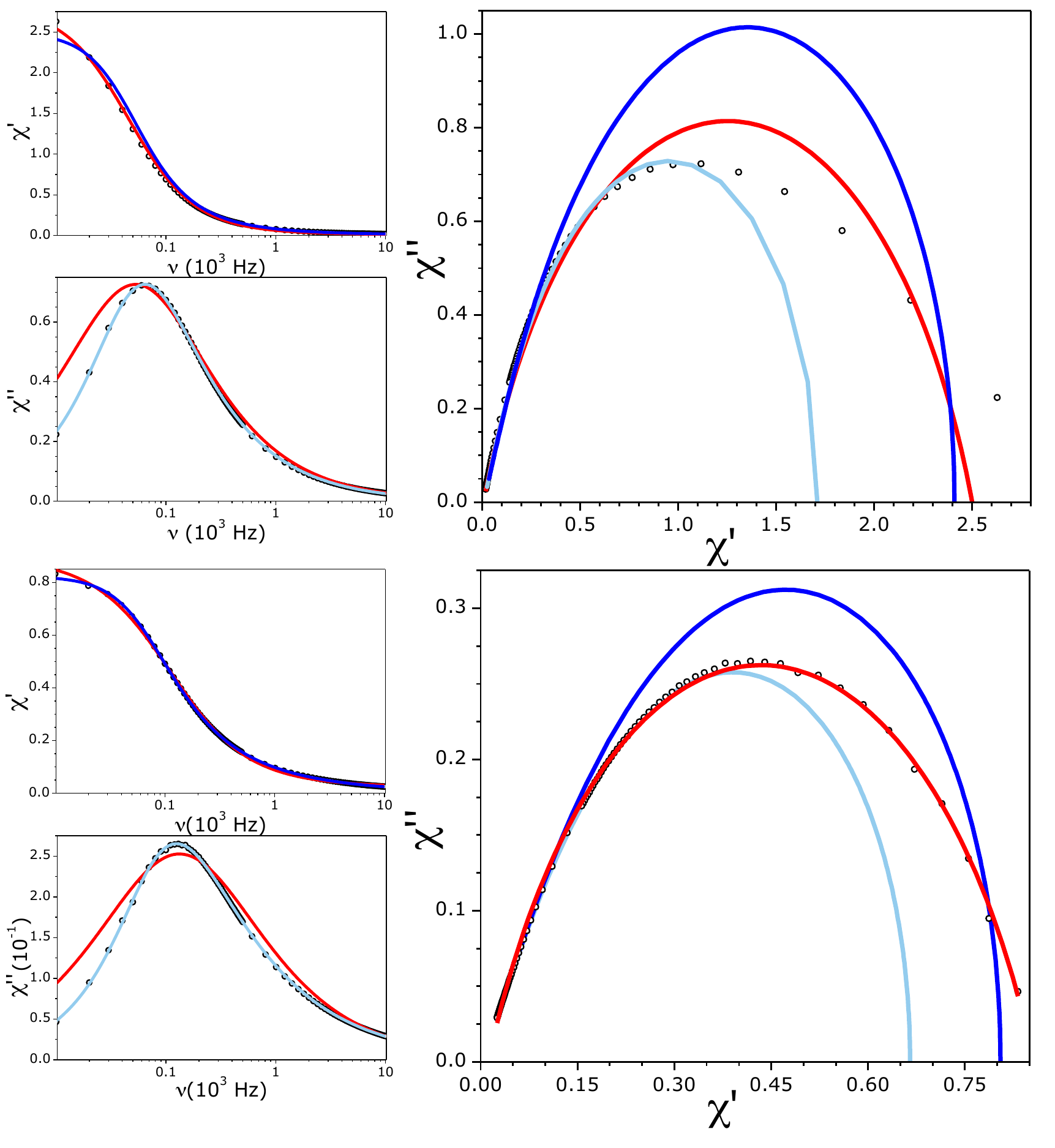} 
\caption{ {\bf Comparison between Cole-Cole (CC) and Cole-Davidson (CD) model. } Top) Data taken at zero dc-field and $T = 1.95$ K: (upper small) real and (lower small) imaginary susceptibility and (main) argand diagram. Bottom) Data taken at zero dc-field and $T = 6$ K: (upper small) real and (lower small) imaginary susceptibility and (main) argand diagram. CC model: The red line fits all the data with the same set of four parameters: $\chi_T,\chi_S, \alpha$ and $\tau$, as defined in the main text. The deviation that happen at low temperature and low frequency is discussed in the text. CD model: The blue and turquoise lines represent the two best fit obtained analysing experimental data. It must be noted that, unlike the CC model, there is not a unique set of parameters that can fit simultaneously the three diagrams. }
\label{fig:ratio}
\end{figure}

Nevertheless, it is possible to fit all data with the same set of four parameters for each temperature (red lines in Fig. S1 show two examples for data taken at $T = 1.95$ K and $6$ K, respectively). On the contrary, when the CD-model is applied, the fit to the experimental data is poorer, as shown in Fig. S1. In particular, it is not possible to find a unique set of parameters to simultaneously fit the real and imaginary frequency dependence of the susceptibility as well as the argand diagram ($\chi''_T = f(\chi'_T)$). It is clear then that the CC formalism seems to give a more representative picture of the dynamics in \DyT. This seems physically reasonable as the CC model assumes a distribution of logarithmic relaxation times that is cut off exponentially at high frequency, while the DC model assumes a power law to high frequency. The latter is not consistent with what is known about the magnetic dynamics in spin ice, where the monopole hop rate would presumably lead to a cut-off at high frequency. Hence the superior performance of the CC model is plausible. Undoubtedly, the determination of the proper distribution of relaxation times for a spin-ice system could in principle lead to a better interpretation of experimental data, especially of the asymmetry showed at low temperatures in the low frequency regime.

\vspace{0.5cm}
{\bf 2. Relaxation Time}

The magnetic relaxation time as a function of temperature of \DyT ~is displayed in Fig. S2. The general dynamic behaviour, in good agreement with previous experimental data\cite{Snyder}, is consistent with the presence of two different regimes. At enough high temperature, above $10$ $K$, the time scale drops dramatically due to a thermally activated process (Orbach-like mechanism) with an energy barrier of few hundreds kelvin compatible with higher energy crystal field levels being populated. This  barrier, $E_a\approx250$ K, is insurmountable below $10$ $K$, and the system enters a quasi-plateau region. A further lowering of the temperature, below $2$ $K$ - outside the temperature range investigated in this work - would have caused a sharp upturn of the relaxation time\cite{Snyder}. Following the work done by Jaubert and Holdsworth\cite{Jaubert}, it is possible to interpret the quasi-plateau region with the presence of a thermal assisted quantum tunnelling through the crystal field barrier: below $10$ $K$ the spins are Ising like and the system can be represented by stochastic single spin dynamics, or in `monopole' language by the creation and diffusion of the topological defects. 

\begin{figure}
\renewcommand{\thefigure}{S\arabic{figure}}
\centering
\includegraphics[width=0.6\columnwidth]{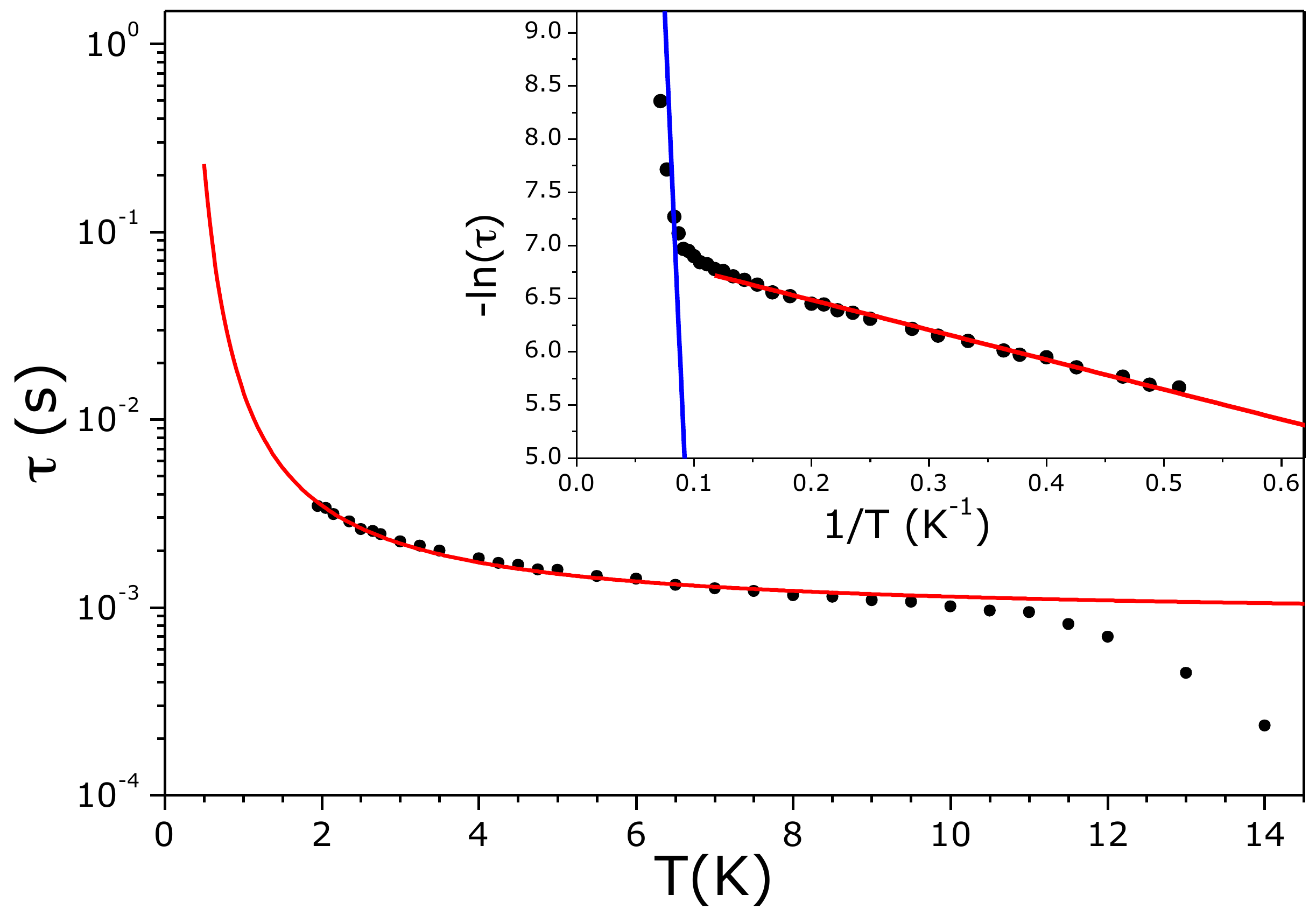} 
\caption{ {\bf Temperature dependence of relaxation time.}  Experimental data extrapolated from the CC-fitting (black circles). The red line is a fit to the free diffusion of topological defects in the nearest neighbour approximation\cite{Jaubert}. Inset: same data shown on a different scale. The blue line represents the fit to an Arrhenius law, with an activation energy of about $250$ $K$, compatible with a classical Orbach-like mechanism (main text).}
\label{fig:ratio}
\end{figure}

In a first approximation~\cite{Jaubert}, it is possible to consider an Arrhenius law of the type:
\begin{equation}
\label{plateau}
 \tau = \tau_0\exp(2J_{eff}/kT),
\end{equation}
where $\tau_0$ is the microscopic tunnelling time and $2J_{eff}$ is the energy cost of a single free topological defect in the nearest neighbour approximation and is half of that of a single spin flip. The red line in Fig. S2 represents the best fit to Eqn. \ref{plateau} with $\tau_0 = 8.6(2)\times 10^{-4}$ $s$ and $2J_{eff}/k = 2.8(1)$ K, in close agreement with the results of Jaubert and Holdsworth~\cite{Jaubert} and in line with the analysis and description that we have proposed in the main paper. 
This expression has been proven~\cite{Jaubert} to work rather well in the semi-plateau regime, but it starts to fail below $2$ K underestimating the relaxation time at very low temperature. Nevertheless, they showed how it is possible to interpret the magnetic relaxation of \DyT, in the whole temperature range in terms of the diffusive motion of monopoles in the canonical ensemble, constrained by a network of `Dirac strings' filling the quasi-particle vacuum.

A similar way to rephrase the problem is to consider that at any given temperature there is only a well defined fraction $x$ of flippable spins (or density of monopoles, reverting to the topological defects representation) that can directly contribute to the magnetic relaxation process. Spins that are not associated with monopoles can flip at much hinger cost and cannot contribute to the dynamics. 
In this work, indeed we have shown that once taken into account the thermal evolution of the density of monopoles and the temperature factor characteristic of a diffusive (Brownian) motion (main text), the resulting characteristic hopping rate turns out to be temperature independent, symptomatic of a spin relaxation that occurs by quantum tunnelling. 

\vspace{0.5cm}
{\bf 3. Ambiguities in the Interpretation of the Dispersion of Relaxation Times}

It should be noted that Eqn. 4 of the main text is only strictly valid for $h_1,h_2 \ll1$ which is marginal for the data considered here. Unfortunately the inversion of a logarithmic time distribution to a frequency distribution is an ill defined mathematical problem~\cite{Zorn} that can only be accomplished in a satisfactory way for narrow distributions. This means that our simple model is not ruled out, but nor is it unambiguously implied by our data. In contrast an unambiguous deduction from our data is that the variance of effective energy barriers is:
\begin{equation}\label{II}
\sigma_E^2 = (kT)^2 (\sigma^2_1 + x \sigma_{2}^2),
\end{equation}
where $\sigma_1,\sigma_{2}$ now become uninterpreted parameters. This quantity is plotted in Fig. S3.

\begin{figure}
\renewcommand{\thefigure}{S\arabic{figure}}
\centering
\includegraphics[width=0.6\columnwidth]{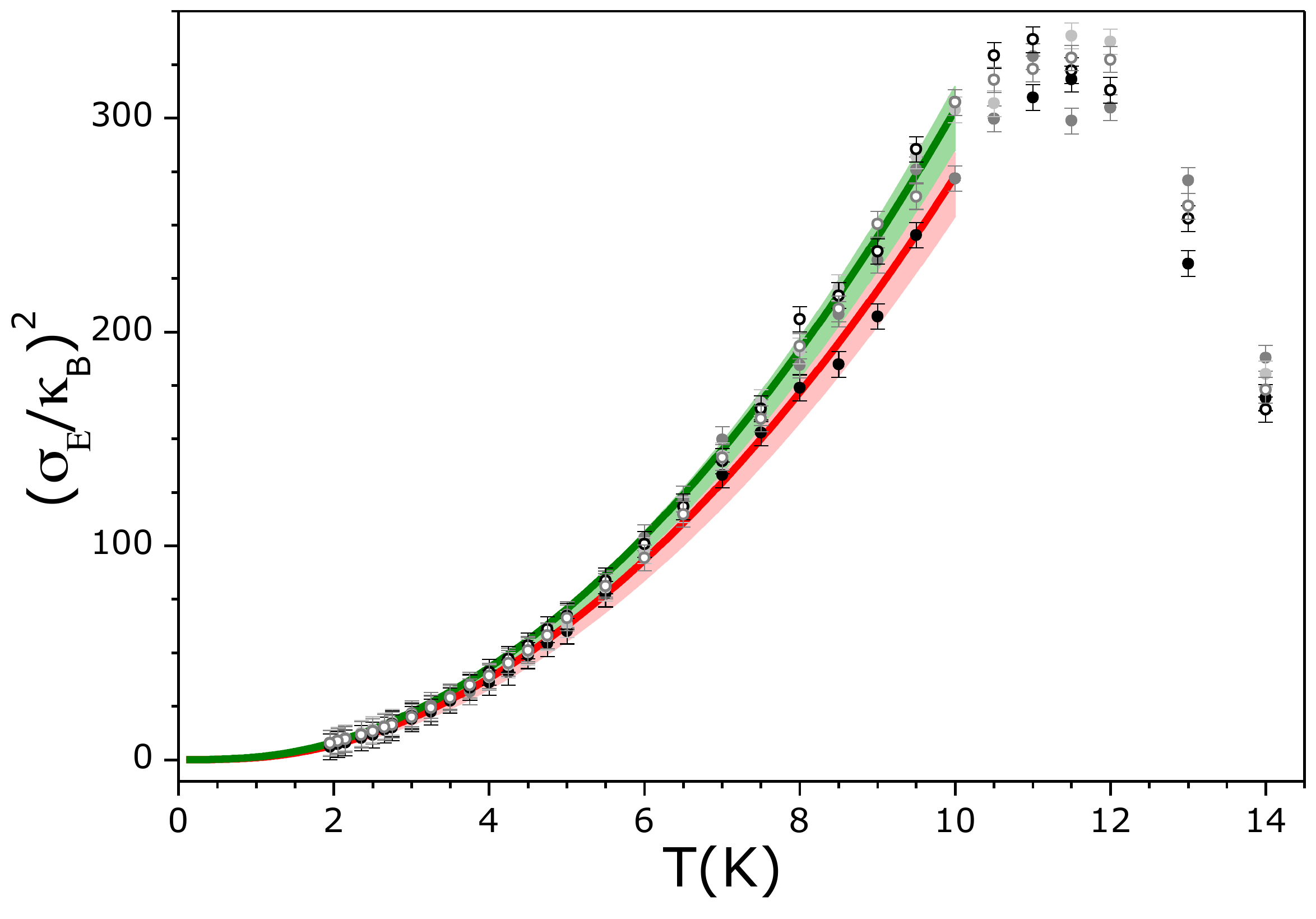} 
\caption{ {\bf Variance of the effective energy activation barrier versus temperature.} Temperature dependence of the experimental variance in energy scale as a function of weak applied field $0\le \mu_0 H \le 0.05$ $T$ (circles, same colour code is maintained as in Fig. 3 of the main text). Red and green indicate, respectively, applied fields of $\mu_0H = 0$ and the set of finite fields. The line is Eqn. \ref{II} with $\sigma_{1,2} = 0.84(1),3.40(5)$ respectively in zero field and $1.15(1),3.40(5)$ in finite field (Fig. 4 main text). The deviation at $T > 10$ K is related to a change in relaxation mechanism (main text). }
\label{fig:ratio}
\end{figure}

While Eqn. \ref{II} is weaker than Eqn. 4 (main text) at the level of physical interpretation, the presence of factors of $kT$ in its right hand side clearly imply a quantum relaxation and its factor of $x$ implies cooperative dynamics involving magnetic monopoles. The direct field interaction that we have suggested is one plausible explanation of this. 

A minimal conclusion that may be drawn from our analysis is as follows. Assuming the monopole model, we can directly measure the diffusion constant $D(T)$. However, this argument cannot be reversed, so the behaviour of the measured relaxation time using the Cole-Cole model does not in itself infer the role of magnetic monopoles. This ambiguity may be traced to the fact that the relaxation due to monopoles takes the form of an effective spin relaxation (see above). In contrast, the fact that the dispersion of the logarithmic relaxation time correlates with the measured monopole density does directly implicate magnetic monopoles in the relaxation process. 

\vspace{0.5cm}
{\bf 4. Isothermal Susceptibility}

\begin{figure}
\renewcommand{\thefigure}{S\arabic{figure}}
\centering
\includegraphics[width=0.6\columnwidth]{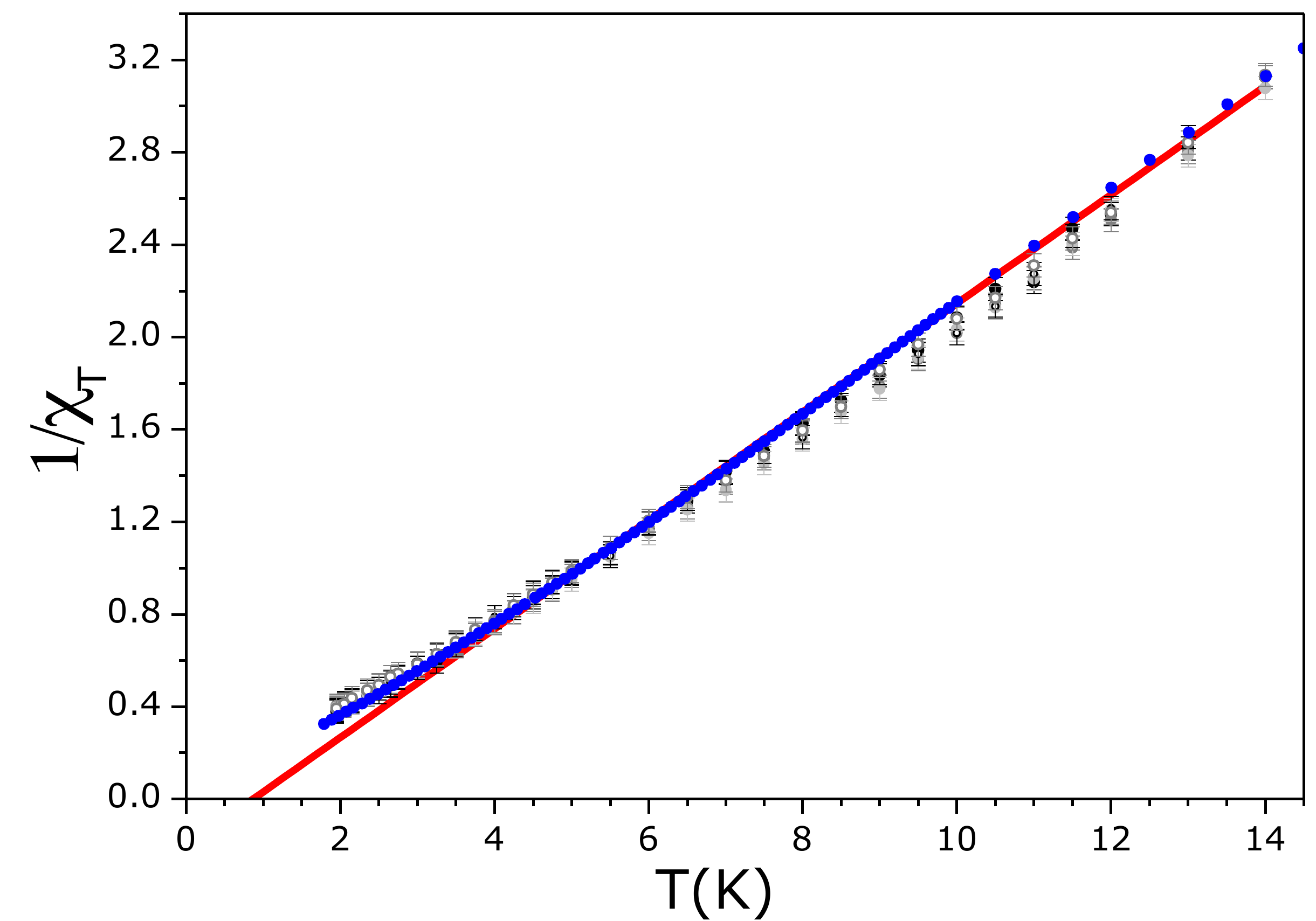} 
\caption{ {\bf Isothermal susceptibility.} Results obtained fitting experimental data in the limit of weak applied field ($0\le \mu_0 H \le 0.05$ $T$); the same colour code is maintained, as in the main paper. The blue circles represent the bulk SQUID magnetometry measurement. The red line is the fit to a Curie-Weiss behaviour as reported in the text.}
\label{fig:ratio}
\end{figure}

For the sake of completeness, Fig. S4 reports the evolution of isothermal susceptibility as a function of temperature in the limit of weak applied field ($\mu_0 H \le 0.05$ $T$). For comparison, the bulk magnetometry SQUID data (blue circles) of the same sample are also displayed in Fig. S4. The best fit (red line) to an apparent Curie-Weiss law $\chi_T$ = $\chi_C/(T-\theta)$ gives rise to a Curie Constant $\chi_C$ = $4.25(5)$ $K$ and $\theta = 0.8(1)$ $K$. Besides, experimental data deviate from this fitting at low temperature: the susceptibility does not approach a constant value as T approaches zero. Similar behaviour was reported for another spin ice compound~\cite{Bramwell,Jaubert-TSF}. As already mentioned in the main text, our results are in good agreement with the prediction~\cite{Jaubert-TSF} that, although the crossover of the Curie constant for a spin ice system is expected to be incomplete in the temperature range studied, the experimental value is higher than the expected $\chi_C$ = $3.95$ $K$ for \DyT.    

\vspace{0.5cm}
{\bf 5. Adiabatic Susceptibility}

To further assess the consistency of our fits, we tested the reliability of the extrapolation of $\chi'(\omega)$ to the limit of $\omega \rightarrow \infty$, to give $\chi_S$. First to exclude the possibility of the presence of an apparent background arising from the instrumental response, we determined the latter by measuring a single crystal of paramagnetic ${\rm LiYF_4}$ doped with $0.3$ $at.$ $\%$ Ho. As shown in the right panels of Fig. S5, the signal is at least two orders of magnitude lower than in the case of the spin-ice crystal in the entire frequency and temperature range. Furthermore, as shown in the left panel of Fig. S5, the temperature dependence of the experimental value of $\chi'$ for ${\rm Dy_2Ti_2O_7}$ measured at the highest applied frequency ($\omega = 10^4$ $s^{-1}$) replicates the temperatures dependence of the estimated $\chi_S(T)$, showing that $\chi'(\omega)$ tends analytically to the high frequency limit.
\begin{figure}
\renewcommand{\thefigure}{S\arabic{figure}}
\centering
\includegraphics[width=0.9\columnwidth]{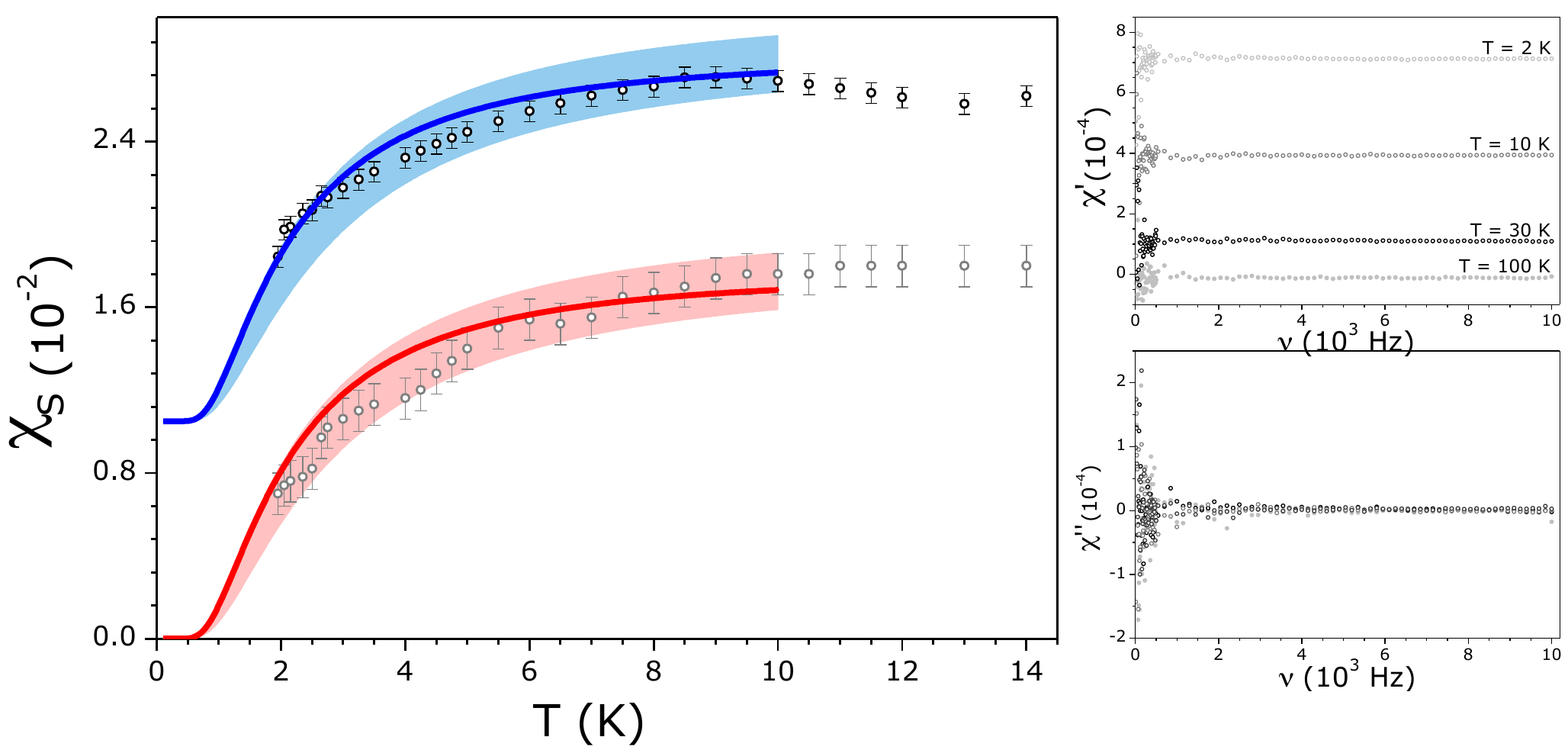} 
\caption{
{\bf Left. Experimental adiabatic susceptibility}: (grey circles) extrapolated from CC-fitting and (black circles) the measured value of $\chi'(10^4~s^{-1}$). The red line represent the expected monopole density adjusted by a scale factor; the shaded area indicates the maximum systematic error in the estimated monopole density (main text and Fig.s 4 and 5). The blue line, and respective shadow, is obtained simply by shifting the red line of an offset of about $0.0105(1)$. {\bf Right. Instrumental response}: Real (upper) and imaginary (lower) component of the susceptibility measured at different temperatures for a single crystal of LiYF$_4$ : $0.3$ $at.$ $\%$ Ho; the signal is at least two orders of magnitude lower than in the case of the spin-ice crystal in the entire temperature range.}
\label{fig:ratio}
\end{figure}

\begin{figure}
\renewcommand{\thefigure}{S\arabic{figure}}
\centering
\includegraphics[width=0.9\columnwidth]{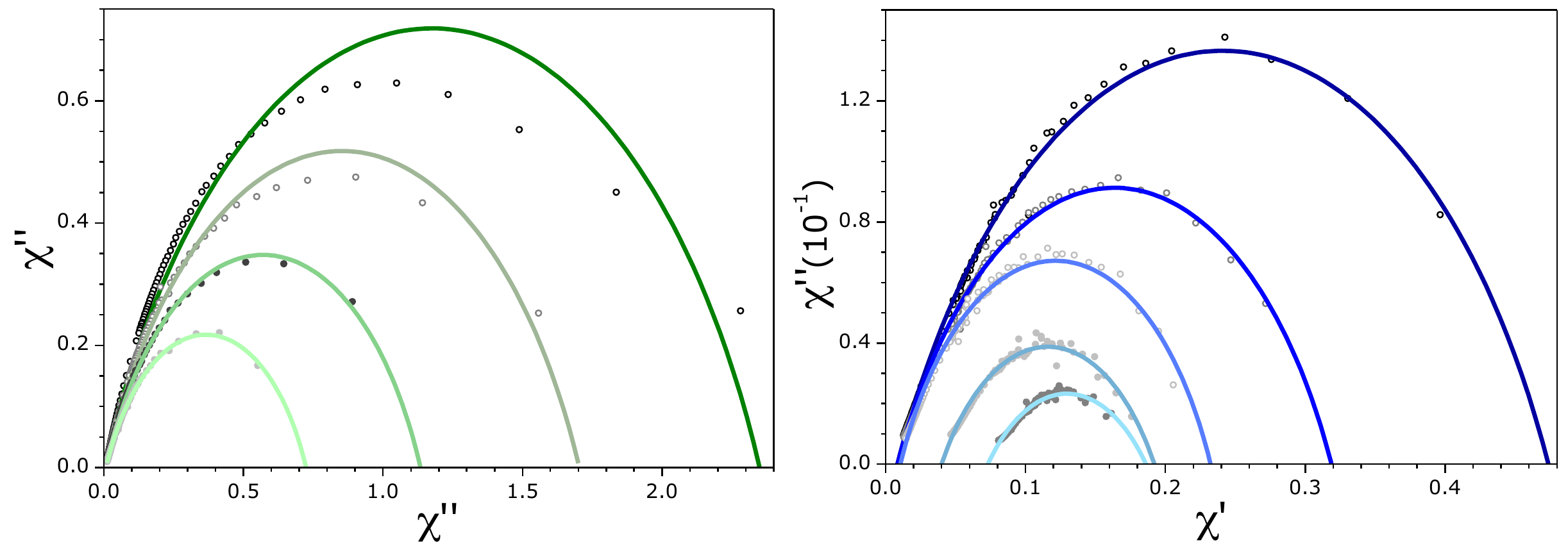} 
\caption{
{\bf Cole-Cole plot and respective CC-fit of data taken at $T = 1.95$ $K$ at different applied field.} Right: representative data taken at, from dark to light green, $\mu_0 H$ of $0.1$, $0.2$, $0.3$ and $0.4$ $T$ respectively. Left: representative data taken at, from dark blue to turquoise, $\mu_0 H$ of $0.5$, $0.7$, $0.8$ and $0.9$ $T$, respectively. }
\label{fig:ratio}
\end{figure}

As described in the main text, we also measured the ac-susceptibility as a function of an applied static magnetic field along [111]; Fig. S6 shows argand diagrams obtained at $T = 1.95$ K. First of all, it is worth notice that the deviation in the low frequency regime of the fitting from the experimental behaviour decreases with the increase of the applied field. Because of the low signal measured at high field, the complete analysis was performed only on data taken at $\mu_0 H < 1$ $T$; at higher field only the frequency dependence of $\chi'(\omega)$ was fitted. For this reason, we preferred to measure the evolution of $\chi_S(T)$ as a function of temperature at $\mu_0 H = 0.86$ $T$ instead of the actual crossover field where $\chi_S$ showed its maximum value (main text and Fig.s 5 and 6).

Looking at the field dependence of the adiabatic susceptibility, according to mean field theory~\cite{Shtyk}, one would expect $\chi_S \sim |H-H_C|^{(-2/3)}$. Fig. 7 shows as an example data taken at $T=1.95K$. The Log-Log plot clearly indicates an exponent of 2 rather than 2/3. 

\begin{figure}
\renewcommand{\thefigure}{S\arabic{figure}}
\centering
\includegraphics[width=0.6\columnwidth]{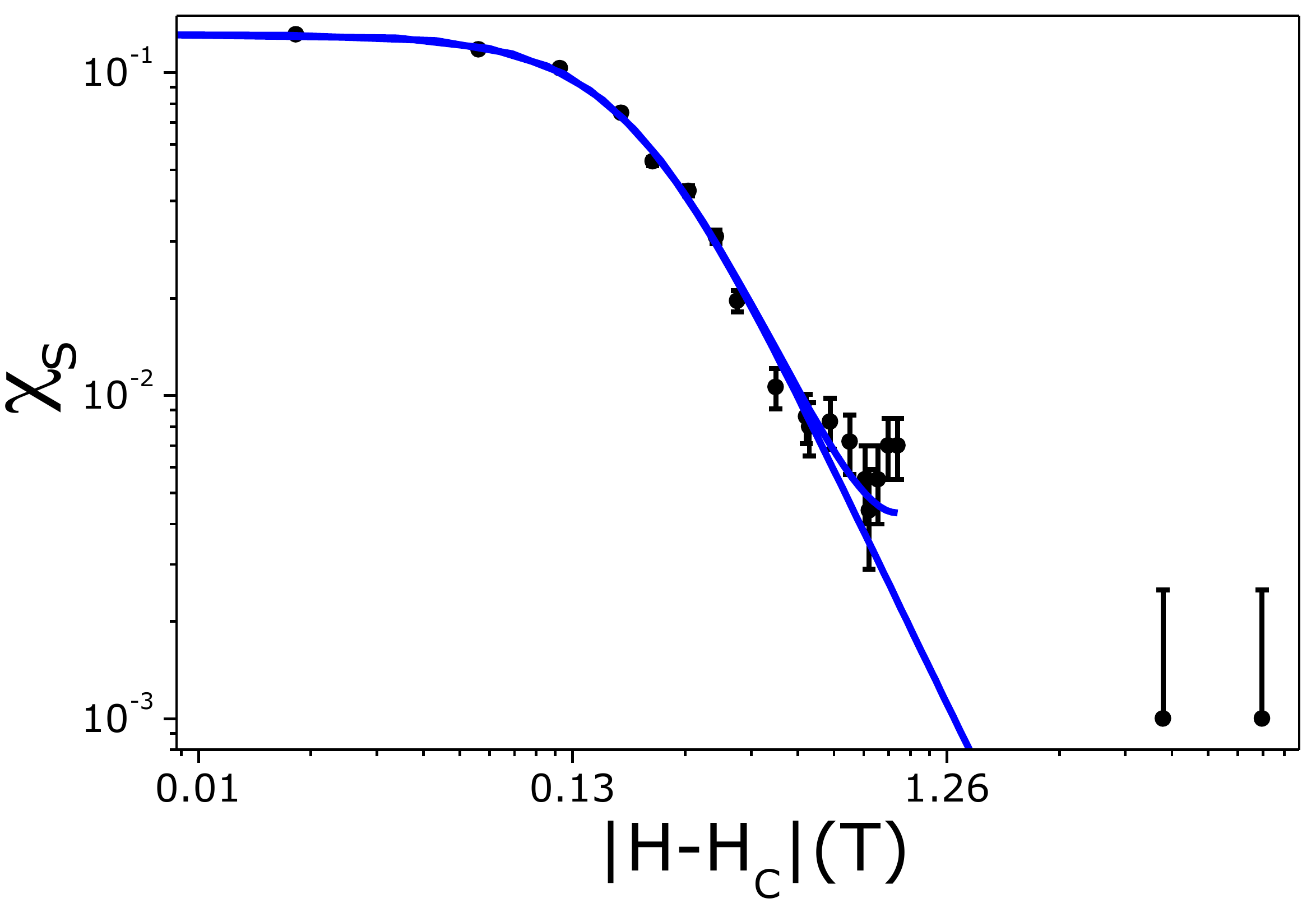} 
\caption{ {\bf Adiabatic susceptibility versus applied field, Log-Log plot.} Experimental $\chi_S$ extrapolated from data at $T =1.95$ K. The blue line shows that the observed exponent of $\alpha = 2$ for $\chi_S \sim |H-H_C|^{(- \alpha)}$.}
\label{fig:ratio}
\end{figure}

\vspace{0.5cm}
{\bf 6. Critical Behaviour}

If we interpret our field dependent measurements in terms of classical critical exponents, then we are led to the unusual conclusion that 
the critical point involved is zero dimensional. Thus, if we assume a zero dimensional critical point ($d=0$) then the scaling laws predict the set of exponents $\delta =-1$, $\gamma=1$, $\beta = -1/2$ and $\nu$ and $\eta$ are of course undefined. Here we 
show how these numbers summarise the experimental observations. 

In critical point theory we can describe the field and temperature dependence of the order parameter $m$ by `polar' co-ordinates $r,\theta$, where $r$ is related to the distance from the critical point in the $\{h,t\}$ plane and $\theta$ is related to the `angle' with respect to the $t$ axis. Here $t = T-T_C$, $h= H-H_C$, $m=M-M_C$. The polar equations in their most general form: 
\begin{equation}
h \sim r^{\beta \delta} f_h(\theta),
\end{equation}
\begin{equation}
t \sim f_t(r,\theta),
\end{equation}
\begin{equation}
m \sim r^\beta f_m(\theta),
\end{equation}
where the three functions $f_{t,m,h}$ on the right hand side are to be determined.

In mean field theory the transformation can be written explicitly: 
\begin{equation}
h \sim r^{\beta \delta} \theta(1-\theta^2)
\end{equation}
\begin{equation}
t \sim r(1-2\theta^2),
\end{equation}
\begin{equation}
m \sim r^\beta \theta,
\end{equation}
with the set of exponents $\gamma = 1, \beta = 1/2, \delta = 3$.

If we write the equations, 
\begin{equation}
h \sim r^{\beta\delta} \tan\theta,
\end{equation}
\begin{equation}
t \sim r,
\end{equation}
\begin{equation}
m \sim r^\beta \theta,
\end{equation}
with the exponents $\gamma=1$, $\beta = -1/2$, $\delta = -1$, we recover our result $\chi = 1/(t+h^2)$. The negative critical exponents look unusual but obey thermodynamics and scaling theory, as is evident from the preceding equations. For example, the Griffiths scaling relation is obeyed}: $\gamma= \beta(\delta -1)$ and $m$ is always an increasing (arctangent) function of $h$. Applying the usual scaling relations we find: $\alpha = 2$, $d=0$ and $\eta,\nu$ are undefined as one would expect for a zero dimensional critical point. As the correlation length and its exponent $\nu$ are undefined, the dynamical exponent $z$ is also undefined. 

Alternatively we could consider a single classical spin in a magnetic field $h$. It is easy to show that to small $h$ the susceptibility is approximated by $\chi \sim T/(h^2+ T^2)$. In this case the polar equations can be satisfied as above, but with $\beta = \delta = 0$; however, the Griffiths scaling relation is not obeyed, suggesting that the classical single spin problem cannot be formally represented as a critical system in the zero temperature limit. 

A final possibility is to interpret the exponents in terms of a quantum critical point. Thus, although the phase transition at zero temperature is first order, at a significant `distance' away the appropriate renormalisation flow might be dominated by a 
$T=0$ critical point of a similar sort to that which occurs the `transverse field Ising model', but which is in reality removed 
by the Coulomb interaction, a strongly relevant variable. A quantum critical point typically exhibits the classical critical exponents appropriate to one higher dimension~\cite{Hertz}. For a three dimensional system this means that quantum critical points are at the upper critical dimensionality of 4, and so exhibit mean field exponents. As shown above, the observed $\gamma$ exponent is mean field like, but when combined with the apparent $\delta$ exponent of $\delta=-1$ suggests a different universality class.  
In the theory of quantum critical points~\cite{Hertz}, this change in dimensionality could be accommodated by a negative dynamical ($z$) exponent, but this would be physically hard to justify. Thus the field dependent behaviour is distinctly anomalous, however it is viewed. 

\vspace{0.5cm}
{\bf 7. Specific Heat}

The specific heat of single crystal ${\rm Dy_2Ti_2O_7}$ was measured with a Quantum Design Physical Properties Measurement (PPMS) System. The Debye-H\"uckel theory of Ref.\cite{CMS2}, along with a correction for double charge monopoles~\cite{Jonathan} was used to fit the specific heat data over the entire range of measured temperature, and so to extract the monopole charge density as a function of temperature. The monopole chemical potential was used as a fitting parameter and was refined to be $\sim 4.33$ $K$, which is in good agreement with theory\cite{CMS2}. The Debye-H\"uckel theory is accurate only for small $x/T$. This means that it is accurate at low temperature, and also a reasonable approximation at high temperature, but breaks down at intermediate temperature, near to the peak in the specific heat at $\sim 1$ $K$. Although this leads to a significant systematic error on the specific heat, the error that propagates through to $x(T)$ is relatively small. Estimates of this systematic error were made by altering parameters in the partition function within reasonable bounds, to give the envelope of curves exhibited in the paper. The estimated $x(T)$ was found to be in close agreement with that derived from numerical simulations \cite{Vojtech} and is clearly sufficiently accurate for our purpose. Full details of the specific heat analysis will be published separately.